# Effect of electron-irradiation on layered quantum materials

Ajit Kumar Dash[1], Mainak Mondal[1], Manvi Verma[1], Keerthana Kumar[1], Akshay Singh[1,*]

[1]Department of Physics, Indian Institute of Science, Bengaluru, Karnataka -560012, India

*Corresponding author: aksy@iisc.ac.in

**Abstract:**

Technological advancement towards the quantum era requires secure communication, quantum computation, and ultra-sensitive sensing capabilities. Layered quantum materials (LQMs) have remarkable optoelectronic and quantum properties that can usher us into the quantum era. Electron microscopy is the tool of choice for measuring these LQMs at an atomic and nanometer scale. On the other hand, electron-irradiation of LQMs can modify various material properties, including the creation of structural defects. We review different types of structural defects, as well as electron elastic- and inelastic-scattering induced processes. Controlled modification of optoelectronic and quantum properties of LQMs using electron-irradiation, including creation of single photon emitters is discussed. Protection of electron-irradiation induced damage of LQMs via encapsulation by other layered materials is encouraged. We finally give insights into challenges and opportunities, including creating novel structures using an electron beam.

**Introduction:**

We are advancing into the quantum age, a step beyond the classical and semi-classical regimes, wherein the quantum nature of information will be used for communication, sensing and computing [1]. The "quantumness" of information plays a key role in each of these separate fields. For example, secure communication can be done by using entangled qubits (quantum bits), wherein the information security is assured by modification of wavefunction by a possible eavesdropper [2]. Sensing capabilities can be extended into the ultra-sensitive regimes by exploiting the effect of external fields on depolarization and decay processes of qubits [3, 4]. On the other hand, quantum computing is a natural consequence of the extended Hilbert space provided by qubits over classical bits [5]. For these exciting quantum information and computing applications, the need of the hour is a reliable way to create isolated single photon emitters (SPEs) as well as array of SPEs, entangled photon pairs, as well as external-stimuli sensitive emitters [3, 6, 7].

Layered materials (LMs) have emerged as a source of these quantum emitters and are broadly classified as layered quantum materials (LQMs) [8, 9]. LQMs vary in a large bandgap range, from 1-6 eV, and include materials like two-dimensional transition metal dichalcogenides (2D-TMDs, example $MoS_2$, $MoSe_2$, $WSe_2$), 2D monochalcogenides (example, GeSe) as well as wide bandgap materials (for example hexagonal boron nitride (hBN)) [10, 11] . It is well known that LQMs have remarkable opto-electronic properties including evolution of bandgap from bulk to 2D, high current on/off ratio, and have light emitting and photonic applications [12–14]. A unique aspect is spin-valley locking which gives rise to the spin-valley index, that can also be used as a qubit [15, 16]. Defects in LQMs are especially interesting because the localized nature of defects can lead to formation of a SPE by cutting down on various decay channels [9]. However, an important challenge in the field is the direct structural measurement as well as deterministic creation of these defects.

Electron microscopy (EM) is a powerful tool to measure the structure at atomic scale and is especially suited for high resolution imaging of LQMs due to extreme thinness of these materials. The thinness of the LQMs allow for direct measurements with the electron beam (e-beam), without repeated beam-material interactions which degrade spatial resolution. EM has been used to measure point defects [17], grain boundaries [18], alloys [19] and oxidation [20, 21] of LQMs. EM has also been used to modify materials at an atomic scale, as well as to create novel geometries [22]. On an application front, e-beam lithography has been used to create sub-20 nm geometries in LQMs [23]. Thus, an understanding of beam-material interactions is very important to fully utilize LQMs for opto-electronic and quantum applications.

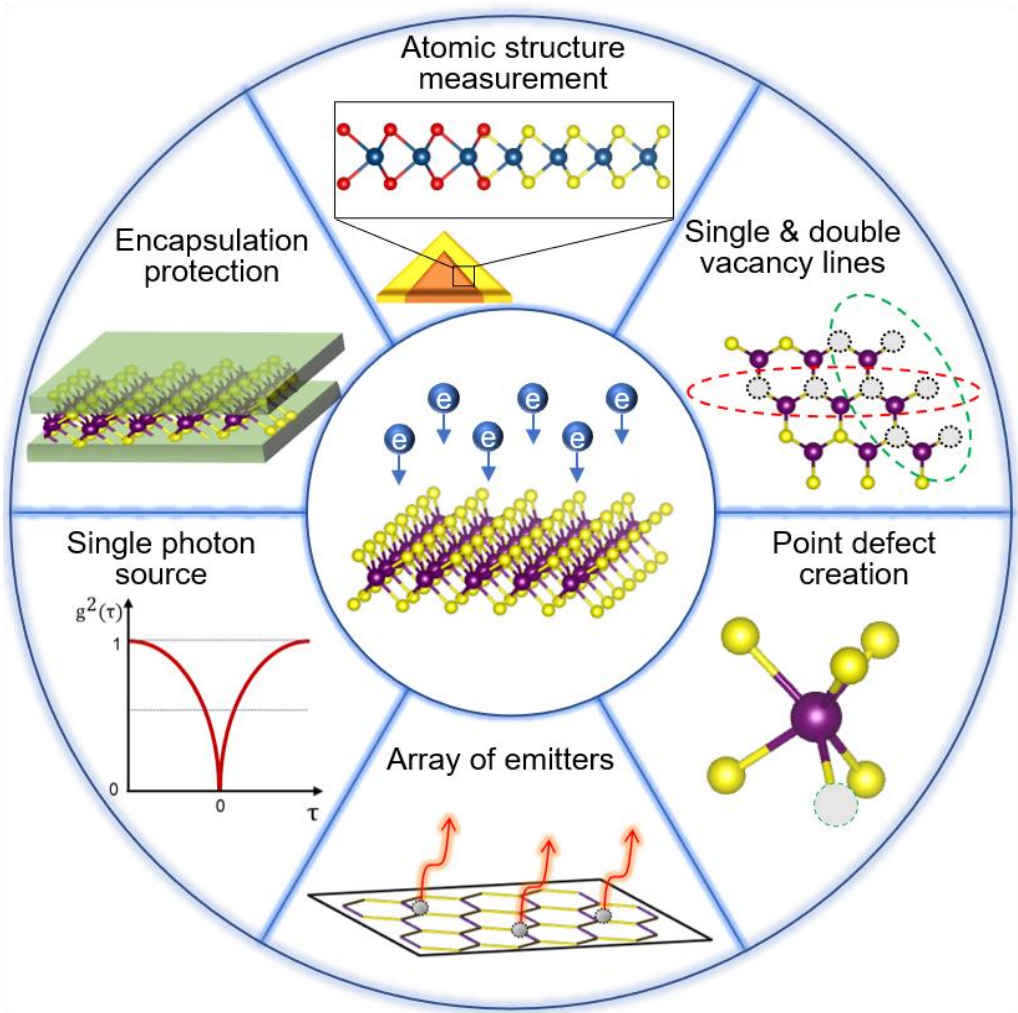

**Figure 1**: Schematic of electron-LQM (layered quantum materials) interactions including measurement of atomic structure, example of irradiation effects and point defect creation. Optoelectronic property modification is illustrated by array of localized emitters and measurement of quantum emission properties. Protection of LQM is possible by encapsulation within other layered materials.

In this review paper, we answer several important questions (Fig 1). How does electron-irradiation affect LQMs. Can we utilize the e-beam to create novel functionalities. While performing structural measurements in a transmission EM (TEM), can we protect the LQMs from electron-irradiation induced damage. We will first discuss the types of defects present in LQMs, followed by electron-irradiation effects on LQMs. We then provide a detailed review of point defects modified by electron-irradiation , as well as the change in opto-electronic and quantum properties. We focus on encapsulation as a viable method to reduce electron-irradiation damage. We finally end with challenges and opportunities in this exciting field.

**Type of defects**

The inevitable presence of defects in LQMs [24] adds to the plethora of interesting and unusual properties possessed by these material [25, 26]. Generally, crystal defects are interruptions of regular patterns in crystalline solids. They are classified, according to their dimensions, into zero, one, two, and three-dimensional defects. However, the defects found in 2D-LQMs are usually zero and one dimensional.

Vacancy defects [17], adatoms [27], substitutional impurities and stone-wales defects are all examples of zero-dimensional (0D) defects [24]. Usually the simplest defects possible [28], the 0D (or point) defects occur at or around a single lattice point (Fig 2(a)). These defects may be intrinsic or extrinsic. Among the former are vacancies (Fig 2 (b),(c)) (lattice sites which would otherwise be occupied in a perfect crystal) and substitutional atoms (Fig 2(d)) (when an atom is substituted with another atom of a different species [29]). Extrinsic defects arise in the presence of an impurity atom which may be interstitial or substitutional. Introducing the impurity atoms might be highly desirable in certain applications (for example, for doping materials) [30, 31].

Dislocations, grain boundaries and edges form important examples of 1D effects usually observed in LQMs. Dislocations are linear defects in which atoms of crystals are misaligned. Grain boundaries (GBs) are present as the interface between two grains with different crystallographic orientation (Fig 2(f)). GBs are usually present in polycrystalline films grown by chemical vapor deposition (CVD) [32], with "tilt" and "mirror-twin" boundaries being the most routinely observed [18, 33]. The scanning TEM (STEM) image of a "mirror-twin" boundary is shown in Fig 2(e). The twin GB is reported to cause strong photoluminescence quenching and enhanced in-plane conductivity [18]. Tilt and mirror twin boundaries might also form in LQMs as a result of chalcogen vacancy evolution [34].

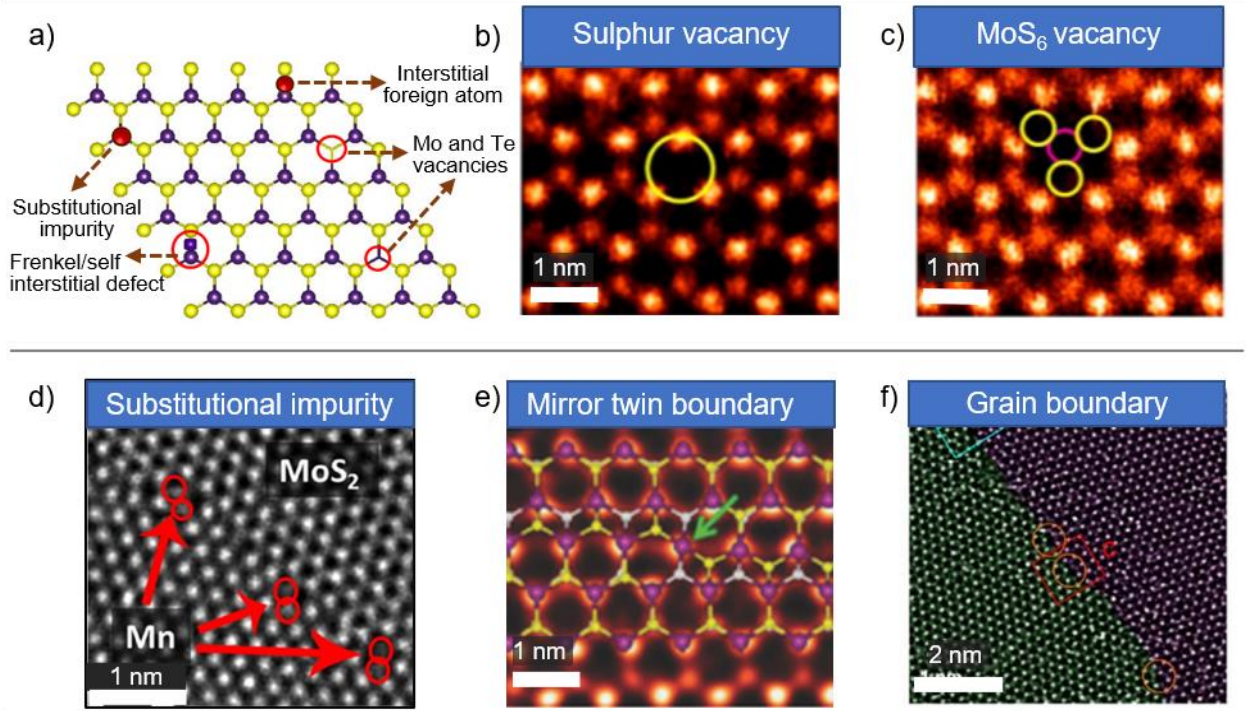

**Figure 2:** **(a)** Schematics of point defects. **(b)** Atomic-resolution annular dark field STEM (ADF-STEM) images of a single sulphur vacancy in CVD-grown monolayer $MoS_2$. **(c)** ADF-STEM images of $MoS_6$ vacancy in CVD-grown monolayer $MoS_2$. (b-c) reprinted with permission from [35]. **(d)** TEM image showing $MoS_2$ monolayers substitutionally doped by Mn atoms. Reprinted with permission from [29]. **(e)** ADF-STEM image of mirror twin boundary in $MoS_2$ [36] **(f)** ADF-STEM image of a grain boundary with 21° tilt angle in single-layered $MoS_2$ film. Common dislocations in $MoS_2$ have been highlighted. Reprinted with permission from [37].

**Electron-irradiation induced changes in LQMs**

Electron-irradiation can create various kinds of structural defects in LQMs. The creation of defects can be due to the displacement of atoms within the specimen or sputtering of atoms from the specimen (Fig 3(a)). The amount of radiation damage and defect creation is proportional to the electron dose [38] . Further, type of defect creation depends on amount of energy transferred during irradiation, which depends directly on the accelerating voltage. High accelerating voltage in a TEM can knock the atoms out from their original position, and create point defects in the sample. The minimum energy transferred by electron-irradiation should exceed the displacement threshold of the atom to be displaced. For example, accelerating voltage (> 80 kV) in a TEM can create nanopores (point defect clusters) in $MoS_2$ by direct sputtering of chalcogen atoms [39].

Defects can also be created by electron-irradiation below the knock-on voltage, with the help of additional energy channels present in the sample. Recent studies show the formation of defects in 2D MoS$_2$ well below the knock-on voltage, with lowering of threshold energy attributed to electronic excitations [40]. Fig 3(b) describes the variation of displacement cross-section of sulphur atoms with accelerating voltage, for different displacement threshold energies. The simulated total cross-section for displacement threshold energy of 1.5 eV is a good fit for the experimental data. The reduction in effective displacement threshold energy of sulphur atoms may be related to contribution from core holes and multielectron excitations. In the excited state, excitations can localize on an emerging defect site, thus lowering the displacement threshold energy. On the other hand, low accelerating voltage (<10 kV) in an SEM was shown to create defects in monolayer MoS$_2$ by hydrocarbon mediated defect formation [41].

Structural defects can evolve under electron-irradiation and develop new morphologies and functionalities. For instance, point defects in mechanically exfoliated monolayer MoS$_2$ under electron-irradiation in a TEM can agglomerate to form single and double vacancy lines [42]. Another example of the evolution of point defects under electron-irradiation is the formation of tilt and mirror twin boundaries in MoSe$_2$ monolayer [34]. These 1D geometries can offer enhanced functionality for optoelectronics and can modify quantum properties by supporting 1D transport. Electron-irradiation can also induce phase transformation among different polymorphs in 2D TMDs. The transition between the semiconducting (2H) phase & metallic (1T) phase in monolayer MoS$_2$ by e-beam irradiation (in a STEM) is one example [43]. Understanding phase transition mechanisms in these LQMs is useful for controlled phase engineering-based applications [12].

In addition, inelastic scattering of incident electrons from the specimen can cause radiolysis, heating, and hydrocarbon contamination. Radiolysis involves the breaking of atomic bonds within the sample in the presence of e-beam flux. Semiconductors and insulators, especially organic compounds, are prone to be damaged by radiolysis [44]. Decreasing the incident beam current lowers the probability of thermal decomposition mediated radiolysis. Radiolysis can also be avoided by cooling the specimen, as radiolysis is temperature-dependent [45]. On the other hand, collision of incoming electrons with electrons of the specimen can cause local heating effects. The temperature rise depends strongly on beam current, accelerating voltage, the thermal conductivity of the material and atomic number [46, 47]. For example, the calculated temperature rise resulting from e-beam incidence (1 keV energy, 100 pA current) on monolayer MoS$_2$ is 0.8°C, and can be neglected [41]. However, at higher accelerating voltages and currents, heating may become a substantial issue. In the specific case of organic and 'soft' materials, beam induced heating is a major concern [48]. Another beam-induced effect is hydrocarbon contamination, in which the carbonaceous contaminants in SEM/TEM column are broken down to free radicals and are deposited onto the sample. These hydrocarbon molecules modify the nature of defects and facilitate the formation of nanopores in single-layer $MoS_2$ [41].

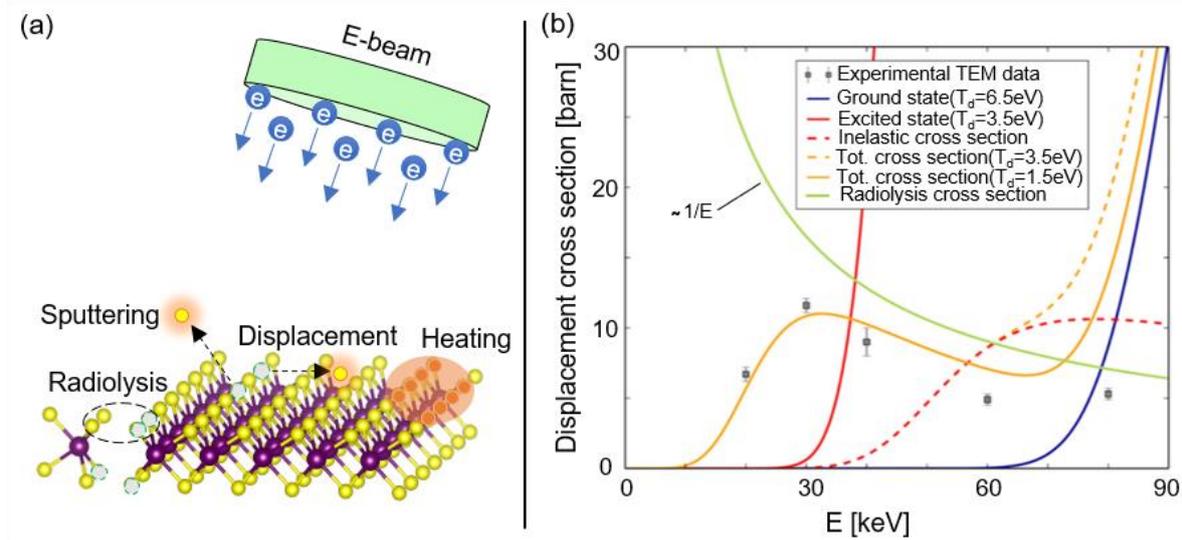

**Figure 3:** Electron-irradiation induced changes in LQMs. **(a)** Schematic of various processes due to electron-irradiation on LQM, including sputtering, specimen heating, atomic displacement, and radiolysis. **(b)** The experimental (squares) and theoretical (curves) displacement cross-sections of sulphur atoms in monolayer $MoS_2$ plotted against accelerating voltage, with different values of displacement threshold $T_d$ corresponding to the ground and excited state. The total cross-section with $T_d = 1.5$ eV matches the experimental data. Reprinted with permission from *[40]*.

**Point defects in TMDs produced by electron-irradiation and their opto-electronic properties**

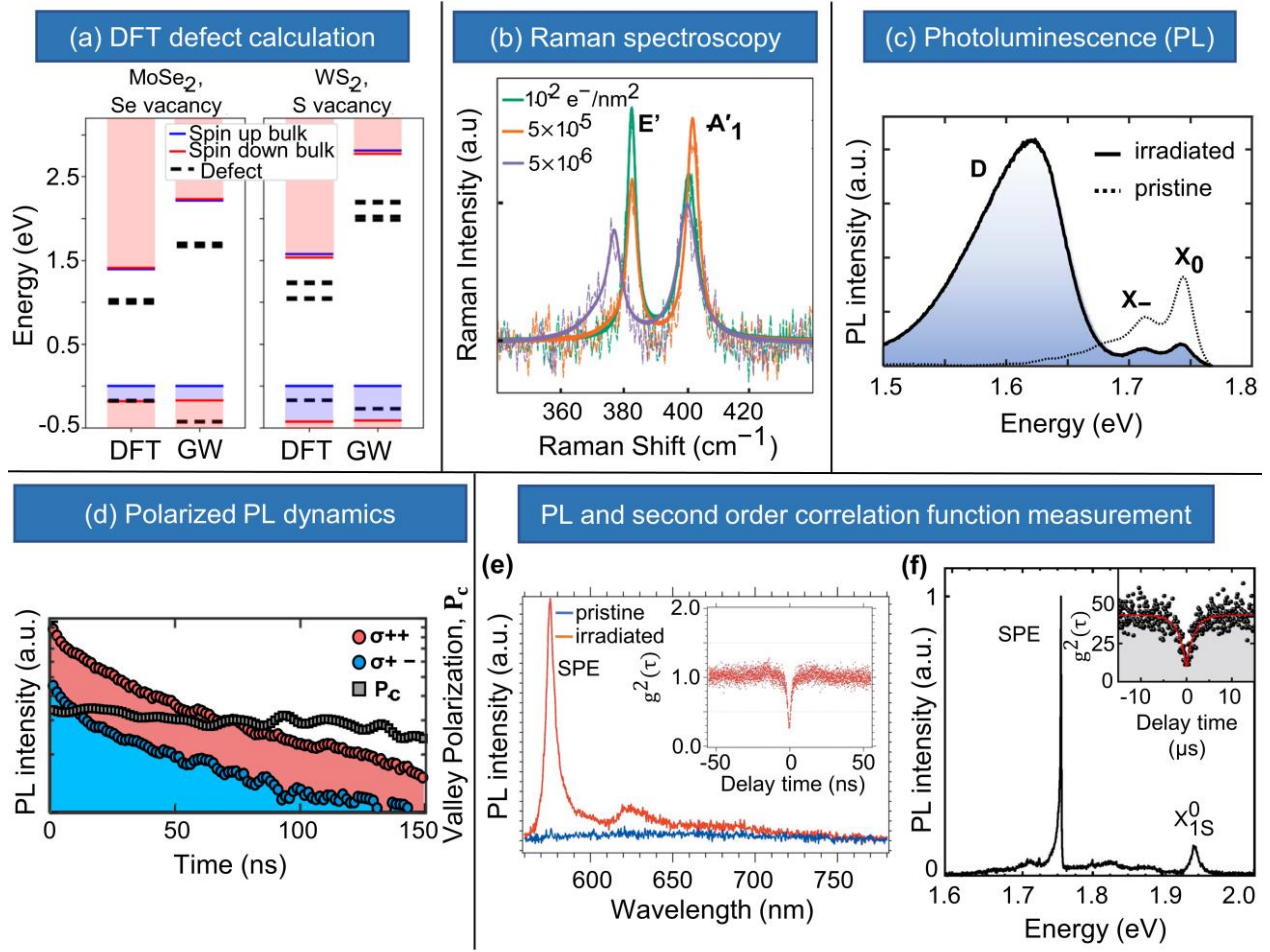

**Figure 4:** Band structure calculation and optical measurements to characterize point defects in LQMs. **(a)** Defect-state energy levels calculated using DFT (density functional theory) and GW for $MoSe_2$ (left) and $WS_2$ (right). Reprinted with permission from [49] **(b)** Raman spectra (dotted lines) and corresponding Lorentzian fits (solid lines) of electron irradiated monolayer $MoS_2$ for low dose region (green), for $5 \times 10^5$ electrons $nm^{-2}$ (orange), and for $5 \times 10^6$ electrons $nm^{-2}$ (purple). Reprinted with permission from [50] **(c)** PL spectrum for pristine (dashed line) and electron irradiated (solid line) monolayer $WSe_2$ at 5 K. For pristine $WSe_2$, neutral ($X_0$) and negatively charged (X-) excitonic peaks are present. After irradiation, a broad spectral band (D) appears at lower energy due to emission associated with selenium vacancies [51] **(d)** Normalized co-circular (red circles) and cross-circular (blue circles) polarized dynamics and the degree of circular polarization ($P_c$) (squares). (c-d) reprinted with permission from [51] **(e)** PL spectrum on few-layer hBN before (blue curve) and after (red curve) electron-irradiation. Inset: second-order correlation function ($g^2(\tau)$) curve confirming that the formed defects act as a single-photon emitter (SPE). Reprinted with permission from [52] **(f)** Low temperature PL of a defect that acts as a SPE in hBN encapsulated monolayer $MoS_2$. The luminescence of the SPE dominates the overall emission spectrum at low excitation powers. Inset: $g^{(2)}(\tau)$ of the PL stemming from one SPE. Reprinted with permission from [53].

According to density functional theory (DFT) calculations, presence of point defects (single or dual chalcogen vacancies) can significantly modify the electronic and phonon band structure. As shown in Fig 4(a), two new energy levels are observed below conduction band minimum (CBM) [49]. The dynamics of these defects will be discussed later. The presence of defects can activate phonon modes situated at the Brillouin zone edge [54–56], whereas in pristine system, vibrational modes only at the $\Gamma$ point are observable. These new modes can shift in-plane and out-of-plane (E' and A'$_1$) vibrational peak positions, as well as modify Raman intensities, as shown in dose dependent measurements on monolayer $MoS_2$ (Fig 4(b)) [50]. Complementary low temperature PL measurements (Fig 4(c)) of an e-beam irradiated monolayer $WSe_2$ (30 keV accelerating voltage) revealed formation of a new broad defect peak (D) [51]. This new peak is ~ 150 meV below and has higher PL intensity compared to neutral ($X_0$) and charged (X-) exciton PL peaks in pristine samples. The D peak can be due to the newly formed defect bands below CBM, as predicted by DFT calculations [49].

In pristine LQMs, bright excitons suffer multiple scattering processes resulting in fast decay rate and a low valley lifetime. However, in a defect embedded system, bright excitons can get trapped in localizing potential wells formed around point defects [16, 57, 58]. Due to localization in these potentials, defect bound excitons are partially protected from valley scattering events. Circularly polarized time-resolved PL measurements for defect states indicate (Fig 4(d)) carrier and valley scattering lifetime to be 200 ns and ~ 1 $\mu$s respectively [51]. This is a large increase over pristine monolayer $WSe_2$, where the recombination and valley lifetime is in the picosecond range [15, 16, 59].

Point defects in LQMs show analogous behavior to 0D quantum dots. The spatially localized excitons in these defects regions show extremely low linewidth optical emissions [60–63]. Second order correlation measurements ($g^2(\tau)$) show strong photon antibunching, which show that these defects can serve as single photon emitters (SPE) [60, 61, 63, 64]. In 2016, two groups Tran et al.[65] and Choi et al. [52] reported that SPEs in hexagonal boron nitride (hBN) can be produced by electron-irradiation and high-temperature annealing. A comparison of pristine and irradiated few-layer hBN samples show much higher intensity PL peaks in irradiated samples, and $g^2(0) < 0.5$ supports the assignment of peak to SPE (Fig 4(e)) [52].

In contrast to e-beam irradiation, a He-Ion beam (30 keV) is expected to produce defects with large formation energies, that are rare in pristine samples (for example, Mo vacancy) [66]. By irradiating with a nanometer-scale He-Ion beam, Barthelmi et al. [53] showed site-selective and deterministic creation of point defects in a hBN encapsulated monolayer $MoS_2$. Low temperature PL and $g^2(\tau)$ measurements performed on these defect sites show single photon emissions with microsecond order recombination time (Fig 4(f)).

Apart from irradiation techniques, strain engineering is also a very promising technique for creating SPEs. Branny et al. [67] and Palacios-Berraquero et al. [68] independently reported that an array of strain defects can be produced by placing pristine monolayer $WSe_2$ or $WS_2$ flake on a nano-pillar substrate (using e-beam lithography). These strained sites show efficient quantum emissions at cryogenic temperatures. In a more recent study, Parto et al. [69] use hBN encapsulated monolayer $WSe_2$ on a nano-pillar substrate and use electron-irradiation on the strained regions. Single photon emissions up to 150K was observed in this work. These studies prove the effectiveness of combining defect engineering with strain engineering to produce quantum emitters.

**Encapsulation and other beam barriers for LQM protection**

As we discussed in the previous sections, electron-irradiation has emerged as an increasingly useful and adaptable technique [70] to measure and modify LQMs [71, 72]. However, every technique has its limitations and challenges. Due to exposed surfaces, LQMs are very sensitive to damage by surface adsorbates, optical or electronic interactions [73]. The e-beam can deliver excess energy to the sample, resulting in structural changes [74] and radiation damage. This puts a limit on the possible spatial resolution and consequently, the obtainable structure and opto-electronic information. To protect the LQM pristine samples, several routes have been applied varying in their degree of applicability and effectiveness. Lowering the administered electron dose is an obvious way to minimize degradation. However, lowering the dose reduces the signal-to-noise (S/N) ratio and limits ability to image LQMs [75].

A powerful method which evolves naturally from the layered nature of LQMs is encapsulation with other LMs. There are three primary advantages, first is that the encapsulating layer acts as a 'sacrificial layer' which is destroyed/sputtered before the sample gets damaged [48]. Second advantage is the reduction of escape of "knock-on" atoms into vacuum, while promoting the damage-reversing recombination reactions at the same time. The third advantage is only applicable for conducting encapsulating layers, wherein electrostatic charging is reduced.

Graphene is particularly suitable for encapsulation as it produces comparatively low background signals (while performing electron microscopy) and aids with increased stability by minimisation of charging effects and external vibrations. Moreover, its crystalline structure can be easily subtracted from (S)TEM micrographs by Fourier filtering. For example, damage in $MoS_2$ monolayer by a 60 keV e-beam can be prevented if the sample is encapsulated by graphene. This three-stack arrangement allows for delivery of doses as high as $1.7 \times 10^{11}$ e/A$^2$ without any damage [76]. Encapsulation is especially useful for 2D-LQMs since an atom (in an un-encapsulated LQM) once "knocked-out" has higher chances of escaping to vacuum, owing to lower number of neighbouring atoms [48]. Graphene liquid cells have also proved to be effective for liquid encapsulation in liquid-phase TEM [77], in place of conventional silicon nitride cells. Graphene acts as a sink for the excess charge, owing to its excellent conductivity, thus, inhibiting the hydroxyl radical formation and subsequent damage to sensitive specimens.

On the other hand, hBN is not only a useful LQM, but is also used for encapsulating TMD monolayers for preparing high quality electronic devices [78, 79]. Such buried interfaces can be measured using cathodoluminescence (e-beam induced luminescence), wherein hBN serves dual purposes. One is that hBN provides a large beam interaction volume, increasing CL yield [75, 80]. Second is the reduction of beam damage [73, 80], where the encapsulation acts as a protective barrier against chalcogen vacancy formation by preventing the escape of free chalcogen atoms [75]. The encapsulation thus allows high S/N and reduced damage CL measurements.

ALD (atomic layer deposited) HfO$_2$ has proven to be an excellent material for encapsulation in the context of electromagnetic radiation [81]. Particularly, MoS$_2$ photodetectors encapsulated by ALD-HfO$_2$ show improved device performance, significant n-type doping and removal of positive charge from the surface. Encapsulation by similar heavy metals/metal-oxides might be extended to the case of LQM protection from electron-irradiation induced damage through further studies.

**Challenges and opportunities**

Even though great progress has been made in the field of electron-LQM interactions, there are many outstanding questions, which naturally lead to opportunities. Since the defect potential drastically changes the quantum properties, a way to engineer the depth of defect potentials via electron-irradiation is needed. A higher defect potential can lead to SPE applications at higher temperatures (for example room temperature). Additionally, precise control on defect placement is necessary for fully utilizing these defects, thus offering an opportunity to utilize irradiation-based techniques. The effect of moire potentials on defects is in the nascent stages, and promises to provide another tuning knob [82]. Theoretical studies to predict defect energy levels and lifetimes will help for efficient search for SPE hosting LQMs.

Regarding beam interactions with LQMs, the exact role and benefit of encapsulation in the cases where dissociative excitation of molecules by inelastic collisions with the beam takes place, needs to be established through further studies [44, 48]. As the decoupling of different damage modes is extremely difficult to be realised practically, molecular dynamic and Monte-Carlo simulations [48] will be very useful. For example, electron-irradiation on hBN nanostructures has been studied using Monte-Carlo simulations, and can be improved by taking in account collision cascades [83, 84]. Currently, the most widely used LMs for encapsulation purposes are graphene and hBN. Using other LMs for the same purpose is still a challenge owing to the lack of control over the modulation of structural and electronic properties of the encapsulated LQM [85].

E-beam is increasingly being used as a nano-sculpting tool, with precise atomic scale placement of atoms, and creating 1D geometries. For example, fabrication of flexible metallic nanowires of TMDs has been carried out by ionization etching in a TEM [86]. Single atom motion can also be induced using a focused e-beam in a STEM, which is now being explored to create atomic scale geometries which are not possible with any other fabrication method [22]. These methods offer a way to create quantum relevant defects and geometries.

## Acknowledgements

A.S. would like to acknowledge funding from Indian Institute of Science start-up grant.

## Author Contributions

All authors contributed to writing of the review paper.

## Competing Interests statement

The authors declare no competing interests.